\documentclass[conference]{IEEEtran}
  \usepackage[left=0.62in, right=0.63in, bottom=0.96in, top=0.7in]{geometry} 
\usepackage{url}
\usepackage[utf8]{inputenc} 
\usepackage[ruled,vlined]{algorithm2e}
\usepackage{amsmath}
\usepackage{amsfonts}
\usepackage{amssymb}
\usepackage{amsthm}
\usepackage{dsfont}
\usepackage{array}
\usepackage{multirow}
\usepackage{lipsum}
\usepackage{mathtools}
\usepackage{cuted}
\usepackage[caption=false]{subfig}
\usepackage{bbm}
\usepackage{siunitx}
\usepackage{bm}
\usepackage{booktabs}
\usepackage{diagbox}

\usepackage{xcolor}

\usepackage{stfloats}
\usepackage{rotating} 
\usepackage{cases}

\usepackage{cleveref}
\usepackage{lineno}
\DeclareMathOperator{\EX}{\mathbb{E}}

\newcommand{\setN}{\mathcal{N}}
\newcommand{\setS}{\mathcal{S}}

\newcommand{\setK}{\mathcal{K}}
\newcommand{\setM}{\mathcal{M}}

\newcommand{\setD}{\mathcal{D}}

\newcommand{\setC}{\mathcal{C}}

\newcommand{\bp}{\bm{p}}

\newcommand{\bw}{\vec{w}}

\newcommand{\bg}{\vec{g}}
\newcommand{\by}{\vec{y}}
\newcommand{\bn}{\vec{n}}

\DeclareSIUnit \belm {Bm}
\newcommand{\numdBm}[1]{\SI{#1}{\deci\belm}} 

\newcommand{\R}{\ensuremath{\mathbb{R}}}  
\newcommand{\C}{\ensuremath{\mathbb{C}}}  

\renewcommand{\vec}[1]{\ensuremath{\bm{\MakeLowercase{#1}}}}

\usepackage{dsfont}

\newcommand{\defeq}{\ensuremath{\triangleq}} 

\Crefname{equation}{Eq.}{Eqs.}
\Crefname{figure}{Fig.}{Figs.}

\usepackage{enumitem}

\usepackage{acro}

\newtheorem{proposition}{Proposition}

\usepackage[final, authormarkup=none, todonotes={textsize=tiny, textwidth= 40pt}]{changes}
\setdeletedmarkup{} 
\definechangesauthor[name={}, color=blue]{all}
\definechangesauthor[name={}, color=orange]{pz}
\newcommand{\pzr}[2]{\replaced[id=pz]{#1}{#2}}

\IEEEoverridecommandlockouts
\IEEEpubid{\begin{minipage}{\textwidth}\ \\[50pt]
		{This work has been submitted to the IEEE for possible publication. Copyright may be transferred without notice, after which this version may no longer be accessible.}
\end{minipage}}

\DeclareAcronym{FL}{
	short = FL,
	long  = federated learning
}

\DeclareAcronym{CR}{
	short = CR,
	long  = communication round
}

\DeclareAcronym{OTA}{
    short = AirComp,
    long  = over-the-air computation
}

\DeclareAcronym{TDMA}{
    short = TDMA,
    long = time-division multiple access
}
\DeclareAcronym{OFDM}{
    short = OFDM,
    long = orthogonal frequency-division multiplexing
}
\DeclareAcronym{MSE}{
    short = MSE,
    long = mean-squared error
}
\DeclareAcronym{SNR}{
    short = SNR,
    long = signal-to-noise ratio
}
\DeclareAcronym{PAPR}{
    short = PAPR,
    long = peak-to-average power ratio
}
\DeclareAcronym{CNN}{
    short = CNN,
    long = convolutional neural network
}
\DeclareAcronym{AWGN}{
    short = AWGN,
    long = additive white Gaussian noise
}
\DeclareAcronym{CSI}{
    short = CSI,
    long = channel state information
}
\DeclareAcronym{IDFT}{
    short = IDFT,
    long = inverse discrete Fourier transform
}
\DeclareAcronym{DFT}{
	short = DFT,
	long = discrete Fourier transform
}
\DeclareAcronym{FFT}{
    short = FFT,
    long = fast Fourier transform
}

\DeclareAcronym{RF}{
	short = RF,
	long = radio-frequency
}
\DeclareAcronym{UE}{
	short = UE,
	long = user equipment
}
\DeclareAcronym{BS}{
	short = BS,
	long = base station
}

\DeclareAcronym{TSE}{
	short = TSE,
	long = true squared error
}

\DeclareAcronym{ICF}{
	short = ICF,
	long = iterative clipping-and-filtering
}

\begin{document}
 \bstctlcite{MyBSTcontrol}
\title{
On Signal Peak Power Constraint of Over-the-Air Federated Learning
\vspace{-.1cm}
}
 \author{\IEEEauthorblockN{Lorenz Bielefeld\IEEEauthorrefmark{2}, Paul Zheng\IEEEauthorrefmark{1}, 
  Oner Hanay\IEEEauthorrefmark{4}, 
  Yao Zhu\IEEEauthorrefmark{3},
 Yulin Hu\IEEEauthorrefmark{3}, and Anke Schmeink\IEEEauthorrefmark{1}}
 \IEEEauthorblockA{\IEEEauthorrefmark{1}Chair of Information Theory and Data Analytics, 
 	RWTH Aachen University, Germany.\\
 Email: $zheng|schmeink$@inda.rwth-aachen.de
 }
 \IEEEauthorblockA{\IEEEauthorrefmark{2}RWTH Aachen University, Germany. 
 Email: $lorenz.bielefeld1$@rwth-aachen.de}
 \IEEEauthorblockA{\IEEEauthorrefmark{3}School of Electronic Information, Wuhan
 	University, China.
 Email: $yao.zhu|yulin.hu$@whu.edu.cn}
  \IEEEauthorblockA{\IEEEauthorrefmark{4}InCirT GmbH, Germany.}
 \vspace{-.9cm}

 \thanks{
 	L. Bielefeld and P. Zheng and co-first authors.
 Y. Zhu and Y. Hu are the corresponding authors.
}
 	  }


\IEEEoverridecommandlockouts

\maketitle

\begin{abstract}
\Ac{FL} has been considered a promising privacy preserving distributed 
edge learning framework. \Ac{OTA} leveraging analog transmission enables the aggregation of local updates directly over-the-air by exploiting the superposition properties of wireless multiple-access channels, thereby alleviating the communication bottleneck issues of \ac{FL} compared with digital transmission schemes. 
This work points out that existing \ac{OTA}-\ac{FL} overlooks a key practical constraint, the instantaneous peak-power constraints due to the non-linearity of radio-frequency power amplifiers. Operating directly in non-linear region causes in-band and out-of-band distortions.
We present and analyze the effect of the \pzr{default}{classic} method that limits the signal's peak power and out-of-band distortions, iterative amplitude clipping combined with filtering.
We investigate the effect of imposing instantaneous peak-power constraints in \ac{OTA}-\ac{FL} for both single-carrier and multi-carrier \ac{OFDM} systems.
Simulation results demonstrate that, in practical settings, the instantaneous transmit power in AirComp-FL regularly exceeds the power-amplifier linearity limit. As the first work of this line of research, it is essential to evaluate if this is an actual problem that has an impact on FL performance. We therefore apply the classic method of iterative clipping and filtering, and show that the FL performance degrades more or less depending on the scenarios. The degradation becomes pronounced especially in multi-carrier OFDM systems due to the in-band distortions caused by clipping and filtering.
\end{abstract}

\IEEEpeerreviewmaketitle

\acresetall

\vspace{-.15cm}
\section{Introduction}
\Ac{FL}, as a distributed edge learning paradigm, addresses privacy and security concerns of deep learning  by keeping raw data on devices and only transmitting model updates for aggregation~\cite{mcmahan_FL_2017}. However, digital transmission of large-scale model updates of massive edge devices creates a communication bottleneck that does not scale well with the number of edge devices~\cite{konecny2017federatedlearningstrategiesimproving, Lim_FL_MEC_Survey_2020}.

\Ac{OTA}~\cite{Nazer_OTA_orig_2007, Slawomir_OTA_robustAnalog_2013} counters this bottleneck in \ac{FL}~\cite{yang_FL_OTA_2020, Amiri_OTA-FL_2020, zhu_OTA-FL_2020} by exploiting waveform superposition of the wireless multiple-access channel to aggregate local updates directly over the air. A central design challenge is power control for amplitude alignment: under average power constraints, threshold-based policies minimizing aggregation \ac{MSE} have been developed, where devices with weak channels transmit at maximum power and stronger-channel devices invert their channels~\cite{Zhang_Tao_2021_GradientStatPowerContFLOTA, Cao_2022_OptimalPowerControlOTAFedAvg, Cao_2022_OptimalPowerControlOTAFL}.

\begin{figure}[t]
    \centering
    \includegraphics[width=0.48\linewidth]{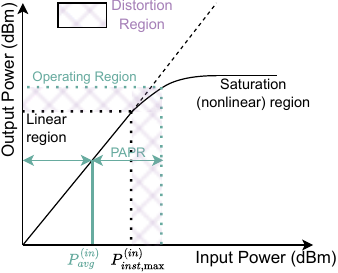}
    \vspace{-.3cm}
    \caption{Typical input and output power characteristics curve for a power amplifier.}
    \label{fig: illustr_PAPR}
     \vspace{-.7cm}
\end{figure}

However, the above designs ignore the instantaneous peak-power limitations due to the non-linearity of power amplifiers as illustrated in Fig.~\ref{fig: illustr_PAPR}. Given a certain average input power $P_{\mathrm{avg}}^{(in)}$, the actual operating region spans up to the maximum peak instantaneous power obtained by multiplying $P_{\mathrm{avg}}^{(in)}$ with the signal \ac{PAPR}. Operating in the nonlinear region causes in-band distortions that corrupt \ac{OTA} aggregation and out-of-band emissions that must be strictly limited, since they may interfere with adjacent frequency bands.
In \ac{OTA}, signal amplitude depends directly on gradient value distributions and may thus encounter high \ac{PAPR}; multi-carrier \ac{OFDM} further inherits its well-known high-\ac{PAPR} problem. While recent works advance \ac{OTA}-\ac{OFDM}~\cite{evgenidis_over--air_2024, Chen_2024_OTA_OFDM, xie_optimal_2024}, they optimize \ac{MSE} under average-power budgets but do not quantify or enforce instantaneous peak-power constraints.

This paper studies \ac{OTA}-\ac{FL} under peak-power constraints for single-carrier and multi-carrier \ac{OFDM} systems. The main contributions are:
\begin{itemize}
    \item We highlight the instantaneous peak-power constraint imposed by power amplifier non-linearity in \ac{OTA}-\ac{FL} for both single-carrier and multi-carrier \ac{OFDM}.
    \item As the first work to highlight this practical issue of high \ac{PAPR} in the context of AirComp-FL, it is essential to confirm that this is an actual issue that can have an impact on FL performance. We therefore propose a detailed characterization of the problem and evaluate the effect of the most classic solution for addressing this issue, \ac{ICF}, with implementation details for both transmission schemes.
    \item Empirical evaluation under LeNet/CIFAR-10: both single-carrier and multi-carrier \ac{OTA}-\ac{FL} exhibit high \ac{PAPR}; clipping degrades accuracy and can cause model divergence.
\end{itemize}

\vspace{-.3cm}
\section{System Model}
\vspace{-.1cm}
\subsection{Federated Learning}
\vspace{-.1cm}

Consider a single-antenna \ac{BS} serving $K$ \acp{UE}, where \ac{FL} is performed. \Ac{UE}~$k$ holds local dataset $\setD_k$ with local empirical loss $F_k(\bw) = \frac{1}{D_k}\sum_{x\in\setD_k} \ell(\bw; x)$, where $\ell(\bw;x)$ is the per-sample loss on~$x$ with model weights $\bw\in\R^N$. \Ac{FL} minimizes $\min_{\bw} \sum_{k\in\setK} D_kF_k(\bw)$, assuming here equal dataset sizes $D_k=D$~\cite{yang_FL_OTA_2020}.
Each \ac{CR} $t$ proceeds as follows:
\begin{enumerate}[leftmargin=14pt]
    \item \Ac{BS} broadcasts global model $\bw^{(t-1)}$ to all \acp{UE}.
    \item Each \ac{UE}~$k$ computes local gradient $\overline{\bg}_k^{(t)}$ via mini-batch SGD on $\setD_k$.
    \item \Acp{UE} send pilots; \ac{BS} estimates channels and feeds back optimal transmit powers (perfect channel state information assumed~\cite{zhu_OTA-FL_2020, yang_FL_OTA_2020}).
    \item  All \acp{UE} apply a shared normalization~\cite{zhu_OTA-FL_2020, zheng_FL-OTA-ISAC_2023} so that $\EX[\bg_k^{(t)}] = 0$, $\EX[\|\bg_k^{(t)}\|^2_2] = N$:
    \vspace{-.2cm}
\begin{equation}
    \bg_k^{(t)} = \frac{\overline{\bg}_k^{(t)} - \mu}{\Gamma},
    \vspace{-.1cm}
\end{equation}
with $\mu^{(t)}=\frac{1}{|\setS_c|}\!\sum_{k\in\setS_c}\!\mu_k^{(t)}$, $\Gamma^{(t)}=\frac{1}{|\setS_c|}\!\sum_{k\in\setS_c}\!\Gamma_k^{(t)}$, where $\mu_k^{(t)}\!=\!\frac{1}{N}\!\sum_i \overline{g}_k^{(t)}[i]$ and $\Gamma_k^{(t)}\!=\!\sqrt{\frac{1}{N}\!\sum_i \overline{g}_k^{(t)2}[i] - \mu_k^{(t)2}}$.
    \item \Acp{UE} transmit $\bg_k^{(t)}$ via analog signal; \ac{BS} recovers the gradient average via AirComp
    \vspace{-.15cm}
\begin{equation}
  \tilde{\bg}^{(t)}\approx  \overline{\bg}^{(t)} \defeq \frac{1}{K}\sum_{k\in\setK} \overline{\bg}_k^{(t)},
    \label{eq: target OTA aggregation}
    \vspace{-.1cm}
\end{equation}
and updates $\bw^{(t)} = \bw^{(t-1)} - \eta \overline{\bg}^{(t)}$.
\end{enumerate}

For ease of notation, the round index $t$ is dropped below.

\vspace{-.1cm}
\subsection{Single-Carrier \ac{OTA}}
\vspace{-.1cm}
\subsubsection{Transmission of gradient}
In the case of single-carrier transmission, each \ac{CR} is divided into $N$ time slots, where $N$ is the size of the trainable parameters of the neural network. In time slot~$n=1,\ldots,N$, all devices transmit their~$n$-th normalized gradient value \(\bg_k[n]\). Through the use of \ac{OTA}, the superimposed characteristics of the wireless medium are exploited to calculate the target function~\eqref{eq: target OTA aggregation}. We assume block-fading, so that the channel characteristics stay constant over the transmission of one gradient vector as assumed in most works~\cite{yang_FL_OTA_2020, zhu_OTA-FL_2020}. The received signal at the centralized server in a given \ac{CR} can be written as 
\vspace{-.2cm}
\begin{equation}
    \begin{aligned}
        \vec{y} =\sum_{k\in\setK} h_k \sqrt{p_k }\vec{g}_k +\vec{n} ,
    \end{aligned} \label{eq:receivedSignal TDMA}
        \vspace{-.1cm}
\end{equation}
with $h_k>0$ the channel gain between BS and \ac{UE}~$k$, $p_k$ the ``average" transmit power of UE~$k$, $\bn\in\R^N$ the additive zero-mean Gaussian noise of variance~$\sigma^2$. 

\Ac{BS} rescales the received superposed signal by a factor $\sqrt{\alpha}>0$ and dividing by the number of UEs~$K$, and then de-normalize the gradient to obtain the $\tilde{\bg}\in\R^N$, the estimation of~\eqref{eq: target OTA aggregation}:
\vspace{-.1cm}
\begin{equation}
    \tilde{\bg} = \frac{\Gamma\sqrt{\alpha}\by}{K } + \mu.
    \label{eq: OTA results}
    \vspace{-.1cm}
\end{equation}

\subsubsection{Power Control}
\Ac{OTA} inevitably faces noises and errors. 
The power $p_k$ and scaling factor $\alpha$ are controlled to minimize the \ac{MSE}
between the AirComp gradient $\tilde{\bg}$~\eqref{eq: OTA results} and the true  gradient $\overline{g}$~\eqref{eq: target OTA aggregation} as 
\begin{equation}
    \begin{aligned}
        \mathrm{MSE} &=\EX[\|\tilde{\bg}-\overline{\bg}\|_2^2]\\
        &= \frac{\Gamma^2}{K^2}\EX[||\sqrt{\alpha }\vec{y} -\sum_{k\in\setK}\vec{g}_k ||_2^2]
        \\[-.2cm]
        &=\frac{\Gamma^2N}{K^2}\Big(\sum_{k\in\setK}(h_k\sqrt{\alpha p_k} - 1)^2+\alpha \sigma^2\Big).
    \end{aligned}\label{eq:MSE_TDMA}
\vspace{-.15cm}
\end{equation}
The derivation uses independence of zero-mean Gaussian noise with~$\bg_k$, $\EX[\|\vec{g}_k\|_2^2]=N$, and $\EX[\bg_k]=0$.

To optimize the MSE, we optimize the term in the outer bracket subject to the average power constraint~\(P_{\mathrm{avg},\max}\). The resulting optimization problem is given by
\begin{equation}
    \begin{aligned}
        \min_{\alpha>0,\ \{p_k\}_{k\in\setK} } \quad &\sum_{k\in\setK}(h_k\sqrt{\alpha p_k} - 1)^2+\alpha \sigma^2\\
        \mathrm{s.t.}\quad\quad \quad& (\forall k\in\setK)\quad 0\leq p_k   \leq P_{\mathrm{avg},\max}.
    \end{aligned}
    \label{pb: single-carrier}
\end{equation}
The optimal solution to the optimization problem has been developed in~\cite{Cao_2020_OptimizedPowerControlOTAFadingChannels}. Later, it was also extended to the case with non-zero-mean gradient, resulting in an additional composite misalignment error in the MSE expression in~\cite{Zhang_Tao_2021_GradientStatPowerContFLOTA}. The optimal solution has a structure of a threshold-based scheme where devices with weaker channels transmit at their maximum average power \(P_{\mathrm{avg},\max}\), while those with channels exceeding the threshold use the respective channel inverting power. 

\begin{proposition}[\cite{Cao_2020_OptimizedPowerControlOTAFadingChannels, Zhang_Tao_2021_GradientStatPowerContFLOTA}]
\label{thm: transmit power single carrier}
Assuming $h_1\geq h_2\geq \ldots \geq h_K$. There exists $k^*\in\setK$ such that the optimal solution $(\alpha^*, (p_k^*)_{k\in\setK})$ to the problem~\eqref{pb: single-carrier} satisfies that the corresponding denoising factor is given by
\vspace{-.3cm}
\begin{equation}
\begin{aligned}
\alpha^* = \frac{1}{P_{\mathrm{avg},\max}}\biggl( \frac{\sum\limits_{k \geq k^*}h_k}{\sum\limits_{k \geq k^*}h_k^2+\frac{\sigma^2}{P_{\mathrm{avg},\max}}} \biggr)^2,
\end{aligned}
\vspace{-.1cm}
\label{eq:optKKT_PowerAllocation_alpha_TDMA}
\end{equation}
and the optimal average transmit power is given by
\vspace{-.2cm}
\begin{equation}
    \begin{aligned}
     (\forall k\in\setK)\quad   p_k^*=
        \begin{cases}
        \frac{1}{\alpha h_k^2}, & \text{if } k<k^*,\\
        P_{\mathrm{avg},\max}, & \text{if } k \geq k^*.
        \end{cases}
    \end{aligned}
    \label{eq:TDMAoptimalPowerControll}
        \vspace{-.1cm}
\end{equation}

\end{proposition}

\subsection{Multi-Carrier OFDM-based \ac{OTA}}
\subsubsection{Transmission of gradient}
In multi-carrier OFDM systems, individual values of \(\vec{g}_k \) are modulated on subcarriers in the frequency domain by assigning one value of \(\vec{g}_k \) to one subcarrier. Since the number of subcarriers \(M\) within the system bandwidth \(B\) is typically less than the size of model parameter~$N$, transmission spans multiple OFDM symbols. 
Let the OFDM symbol index be $\ell\in\setS\defeq\{0,\ldots,L-1\}$ with $L=\lceil \frac{N}{M}\rceil$.
For~$\ell$-th symbol, the subcarrier $m\in\setM \defeq \{0,\ldots, M-1\}$ carries the gradient entry of index $n=\ell M +m$. Zero values are transmitted for $n\geq N$.
The $i$-th time-domain sample of the \(k\)-th \ac{UE}'s \(\ell\)-th OFDM symbol is:
\vspace{-.2cm}
\begin{equation}
   (\forall i\in\setM)\quad     s_{k,\ell}[i] =\sum_{m=0}^{M-1}g_{k}[\ell M +m] b_{k,\ell M +m} e^{j\frac{2\pi mi}{M}},
    \label{eq: multiCarrierTimeDomainSignal}
    \vspace{-.1cm}
\end{equation}
where \(b_{k,\ell M +m}\in\C\) performs channel pre-equalization. Assuming synchronized transmission and cyclic-prefix length exceeding the channel delay spread, applying DFT at the \ac{BS} yields for $m\in\setM$:
\vspace{-.1cm}
\begin{equation}
    Y_{\ell M +m} = \sum_{k=1}^{K}g_{k}[\ell M +m]\, b_{k,\ell M +m}\, H_{k,\ell M +m} + N[m],
    \label{eq:OFDM_received_freqdomain}
    \vspace{-.1cm}
\end{equation}
where \(H_{k,n}=|H_{k,n}|e^{j\phi_{k,n}}\) is the sub-channel response and \(N[m]\sim\setC\setN(0,\sigma^2)\). Setting \(b_{k,n}=\sqrt{p_{k,n}}e^{-j\phi_{k,n}}\), the $n$-th gradient value is recovered as:
\vspace{-.2cm}
\begin{equation}
    \tilde{g}[n] = \frac{\Gamma\sqrt{\alpha_n}}{K}\sum_{k\in\setK}|H_{k,n}|\sqrt{p_{k,n}}\,g_k[n] +\frac{\Gamma\sqrt{\alpha_n}}{K}N[n] + \mu,
    \vspace{-.2cm}
\end{equation}
with $p_{k,n}$ the transmit power of UE~$k$ at subcarrier~$n$ and \(\alpha_n>0\) the denoising factor.
\subsubsection{Power Allocation}
As for single-carrier, the power allocation is done by minimizing the MSE of each gradient value. The MSE of the gradient value~$n$ has the following expression:
\vspace{-.2cm}
\begin{equation}
    \begin{aligned}
        \mathrm{MSE}_n &=\EX[\|\tilde{g}[n]-\overline{g}[n]\|^2]\\
        &=\frac{\Gamma^2}{K^2}\Big(\sum_{k\in\setK}(|H_{k,n}|\sqrt{\alpha_n p_{k,n}} - 1)^2+\alpha_n \sigma^2\Big).
    \end{aligned}           
    \label{eq: MSE_OFDM}
    \vspace{-.2cm}
\end{equation}
For any OFDM-symbol~\(\ell\), the optimal power allocation for gradient values of indices $n=\ell M+m$ with $m=0,\ldots,M-1$ can be given as:
\vspace{-.3cm}
\begin{subequations}
\label{pb:OFDMgenerall Optimization problem}
    \begin{align}
        \min_{\substack{\vec{\alpha}_{\ell} \geq0 \\ (\vec{p}_{\ell M+m})_{m\in\setM} }} &\sum_{m=0}^{M-1}\mathrm{MSE}_{\ell M +m} (\alpha_{\ell M+m} ,\bp_{\ell M+m} ) 
    \\[-.3cm]
        \text{s.t.} \quad\quad\quad& \hspace{-.8cm}(\forall k \in\setK)\quad \sum_{m=0}^{M-1} p_{k,\ell M+m} \leq P_{\mathrm{avg},\max},\label{cons: multi-carrier sum power}\\[-.1cm]
        & \hspace{-.8cm}(\forall k \in\setK)\ (\forall m \in\setM)\ p_{k,\ell M+m}\geq 0,\\[-.6cm]
        & \notag
    \end{align}
\end{subequations}
with $\bm{\alpha}_{\ell} = (\alpha_{\ell M+m})_{m\in\setM}$ and $\vec{p}_n = (p_{k,n})_{k\in\setK}$ for any $n=\ell M+m$.
This problem has been analyzed in~\cite{Chen_2024_OTA_OFDM} that requires a subgradient solver and partly in~\cite{xie_optimal_2024} that considers an equal $\alpha$ for all subcarriers.

Due to the short allocation time and the marginal effect of per-subcarrier power control in uplink OFDM~\cite{Keunyoung_UplinkOFDMA_equalPowerJustification_2005, Yaacoub_LowComplexityOFDMAScheduling_2009}, we assume equal max power per subcarrier, replacing~\eqref{cons: multi-carrier sum power} by $p_{k, \ell M+m}\leq P_{\mathrm{avg},\max}/M$. The problem then decouples per subcarrier~$m$, each solved via Proposition~\ref{thm: transmit power single carrier} with budget $P_{\mathrm{avg},\max}/M$.

\vspace{-.1cm}
\section{Instantaneous peak power constraint}
\vspace{-.1cm}

As presented in the introduction, practical radio-frequency power amplifiers operate linearly only up to certain input power~\cite{Seung_PAPRReduc_2005, Rahmatallah_PAPRReduc_2013}. Operating in linear region would result in in-band and out-of-band distortions.

Note that all previously determined powers are actually output powers from the power amplifier, but can be easily converted to the input power by a one-to-one relationship. We denote $p_k^{(in)}$ and $b_{k\ell M+m}^{(in)}$ respectively the corresponding input power of $p_k$ and $b_{k\ell M+m}$.
 We model this by an instantaneous amplitude limit \(A_{\max}>0\) that corresponds equivalently to the input peak-power limit \(P_{\mathrm{inst},\max}^{(in)}=A_{\max}^2\). The transmitted time-domain waveform \(X(t)\) generated at a \ac{UE} must satisfy
 \vspace{-.1cm}
\begin{equation}
    |X(t)|^2 \leq P_{\mathrm{inst},\max}^{(in)} = A_{\max}^2.
    \label{eq:inst_limit}
    \vspace{-.1cm}
\end{equation}


\subsection{Peak-to-Average Power Ratio}
\vspace{-.05cm}
This limitation directly interacts with the \ac{PAPR} of the transmitted waveform as illustrated in Fig.~\ref{fig: illustr_PAPR}. A low PAPR can keep the operating region within the linear region.

\paragraph{Single-carrier transmission}   With the gradient normalization in Section~II, the average per-\ac{CR} transmit power equals the configured average power budget. The peaks for single carrier transmission are directly determined by the gradient values. The \(k\)-th \ac{UE} transmits the real-valued sample \(g_k[n]\) matching the amplitude of waveform. Its PAPR is
\vspace{-.2cm}
\begin{equation}
    \mathrm{PAPR}_k = \frac{\max_n |g_k[n]|^2}{\frac{1}{N}\sum_{n=1}^N |g_k[n]|^2}.
    \vspace{-.1cm}
\end{equation}

\paragraph{Multi-carrier OFDM transmission} The time-domain signal for multi-carrier OFDM can be written as
\vspace{-.15cm}
\begin{equation}
    s_{k,\ell}(t) =\sum_{m=0}^{M-1}g_{k}[\ell M +m] b_{k,\ell M +m} e^{j\frac{2\pi mt}{M}}.
    \vspace{-.15cm}
\end{equation}
The ideal PAPR is denoted as 
\vspace{-.15cm}
\begin{equation}
    \mathrm{PAPR}_{k,\ell}^{\mathrm{ideal}} = \frac{\max_t |s_{k,\ell}(t)|^2}{\frac{1}{T}\int_0^T |s_{k,\ell}(t)|^2dt}.
\end{equation}
In practice, Nyquist-rate samples underestimate the true PAPR. An oversampling factor $L_{\mathrm{os}}=4$~\cite{Tellabura_PAPR_OversamplingFactor4_2001} is applied by zero-padding the frequency-domain signal ($G_{k,\ell}[m]=g_k[\ell M+m]b_{k,\ell M+m}^{(in)}$ for $0\leq m<M$, zero otherwise) before the IDFT:
\vspace{-.1cm}
\begin{equation}
    s_{k,\ell}^{(\mathrm{os})}[i] = \!\!\!\sum_{m=0}^{L_{\mathrm{os}} M-1}\!\! G_{k,\ell}[m] e^{j\frac{2\pi mi}{L_{\mathrm{os}} M}}, \quad i\!=\!0,\ldots,L_{\mathrm{os}} M\!-\!1.
    \vspace{-.1cm}
\end{equation}
The per-symbol PAPR is then estimated as
\vspace{-.15cm}
\begin{equation}
    \mathrm{PAPR}_{k,\ell} = \frac{\max_i |s_{k,\ell}^{(\mathrm{os})}[i]|^2}{\frac{1}{L_{\mathrm{os}} M}\sum_{i=0}^{L_{\mathrm{os}} M-1} |s_{k,\ell}^{(\mathrm{os})}[i]|^2}.
    \vspace{-.15cm}
\end{equation}

\subsection{Clipping with Oversampling}
The most classic technique without changing the modulation schemes (that are necessary for \ac{OTA}) to deal with the high \ac{PAPR} issue is applying the procedure of \ac{ICF}~\cite{Seung_PAPRReduc_2005}. Since clipping results in in-band and out-of-band distortions in multi-carrier OFDM~\cite{Rahmatallah_PAPRReduc_2013},
a low-pass filter is applied afterwards to limit out-of-band distortion, 
 which may result in the rise of PAPR, necessitating repeating such procedures.

\paragraph*{Amplitude Clipping}
Let \(\mathcal{C}_{A_{\max}}(x) \triangleq \min\{1, A_{\max}/|x|\}\, x\) denote the time-sample-wise amplitude clipper.

\subsubsection{Single-carrier}
Each UE transmits
\begin{equation}
    x_k[n] \!=\! \mathcal{C}_{A_{\max}}\!\bigl(\! \sqrt{\!p_k^{\!(in)}}\!\! g_k[n]\! \bigr) \!=\! \min\Bigl\{1,\! \frac{A_{\max}}{ \sqrt{\!p_k^{\!(in)}}\,\!\!\! g_k[n]} \! \Bigr\}\! \sqrt{p_k^{\!(in)}}\,\!\!\! g_k[n].
    \label{eq:TDMA_clip_correct}
\end{equation}
In single-carrier, this is equivalent to considering the transmission of a clipped version of the gradient values, with nothing else altered by the clipping.

\subsubsection{Multi-Carrier OFDM}

\paragraph{Amplitude Clipping}
For each OFDM symbol \(\ell\), the oversampled complex time-domain samples are clipped while preserving phase:
\begin{equation}
    x_{k,\ell}^{(\mathrm{os})}[i] = \mathcal{C}_{A_{\max}}\bigl( s_{k,\ell}^{(\mathrm{os})}[i] \bigr) = \min\Bigl\{1, \frac{A_{\max}}{|s_{k,\ell}^{(\mathrm{os})}[i]|} \Bigr\} s_{k,\ell}^{(\mathrm{os})}[i].
\end{equation}
Since the clipping operation is nonlinear, distortions are inevitable.
The clipping distortion in the time domain can be written as: $
    d_{k,\ell}^{(\mathrm{os})}[i] = x_{k,\ell}^{(\mathrm{os})}[i] - s_{k,\ell}^{(\mathrm{os})}[i].$

\begin{figure*}
    \centering
    \includegraphics[width=0.65\linewidth]{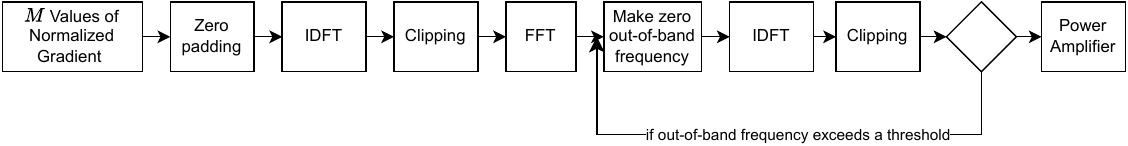}
    \vspace{-.25cm}
    \caption{Flow Diagram for \ac{ICF} in multi-carrier OFDM systems.}
    \label{fig: flow diagram}
     \vspace{-.6cm}
\end{figure*}
\paragraph{Frequency-Domain Distortion}
Transforming the clipped signal back to the frequency domain via DFT yields:
\begin{equation}
    X_{k,\ell}[m] = \mathrm{DFT}\bigl\{ x_{k,\ell}^{(\mathrm{os})}[i] \bigr\}[m] = G_{k,\ell}[m] + D_{k,\ell}[m],
\end{equation}
where \(D_{k,\ell}[m] = \mathrm{DFT}\bigl\{ d_{k,\ell}^{(\mathrm{os})}[i] \bigr\}[m]\) represents the clipping distortion in the frequency domain. 

This distortion consists of two components:
\begin{itemize}[leftmargin=13pt]
    \item \textbf{In-band distortion}: distortion on the considered subcarriers $m\in\setM$, which degrades the aggregated gradients.
    \item \textbf{Out-of-band emissions}: spectral regrowth on the zero-padded subcarriers \(M \leq m < L_{\mathrm{os}} M\) due to clipping, which is to be strictly limited since it may cause interference to adjacent frequency bands.
\end{itemize}
\paragraph{Filtering}
To suppress out-of-band emissions, a rectangular low-pass filter ($H_{\mathrm{filter}}[m]=1$ for $0\leq m<M$, else $0$) is applied in the frequency domain and converted back via IDFT:
\vspace{-.1cm}
\begin{equation}
    x_{k,\ell}^{(\mathrm{filt})}[i] = \mathrm{IDFT}\bigl\{ X_{k,\ell}[m] \cdot H_{\mathrm{filter}}[m] \bigr\}[i].
    \vspace{-.1cm}
\end{equation}
However, filtering may reintroduce peak regrowth~\cite{Seung_PAPRReduc_2005}.

\paragraph{Iterative Clipping and Filtering}
To jointly control PAPR and out-of-band emissions, an \ac{ICF} procedure is commonly employed~\cite{Armstrong_PAPR_repeatedClippingFrequencyDomainFiltering}:
\begin{enumerate}[leftmargin=14pt]
    \item Set \(x_{k,\ell}^{(0)}[i] = s_{k,\ell}^{(\mathrm{os})}[i]\).
    \item Repeat until out-of-band distortion is below threshold: clip $x^{(j-1)}$ to $A_{\max}$; apply DFT; apply $H_{\mathrm{filter}}$; apply IDFT to obtain $x^{(j)}$.
    \item Output \(x_{k,\ell}^{(\mathrm{ICF})}[i] = x_{k,\ell}^{(J_{\max})}[i]\).
\end{enumerate}
The flow diagram is shown in Fig.~\ref{fig: flow diagram}.

\section{Simulations}
\vspace{-.1cm}

\subsection{Simulation Settings}
\vspace{-.1cm}
In the simulations the LeNet convolutional neural network with $62006$ trainable parameters is trained on the Cifar-10 dataset for~$500$ \acp{CR}, where the data is independent and identically distributed among~$K=40$ \acp{UE}. 
In each \ac{CR} all \acp{UE} participate and are uniformly distributed on a~$100$ meter radius disk around the base station. As a large-scale fading model, the free-space-path loss is used, which is combined with Rayleigh fading as a small-scale fading model. The constraint of the maximum average power $P_{\mathrm{avg},\max}$ of \acp{UE} is chosen to be \numdBm{23}, while the constraint of the the maximum instantaneous power $P_{\mathrm{inst},\max}$ is chosen to be \numdBm{26}. In multi-carrier OFDM transmission, $M=32$ subcarriers are used in order to reduce OFDM complexity, where each subcarrier transmits with a bandwidth of \SI{60}{\kilo\hertz}. 
Clipping and filtering iterate until out-of-band emissions are below \numdBm{-10}.
The learning rate is set at 1 and the local batch size at 256.

\subsection{Simulation Results and Discussions}
\vspace{-.15cm}
All curves in this section are smoothed for readability.

\begin{figure}[t]
    \centering
    \subfloat[Single-carrier.\label{fig: tdma_papr}]{
\includegraphics[width=0.49\linewidth, trim=0 0 -10 0]{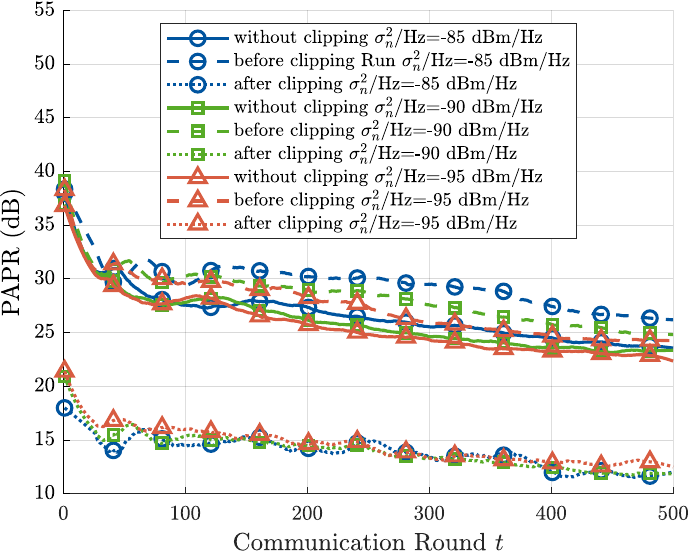}
    }
 \subfloat[Multi-carrier.\label{fig: ordma_papr}]{
\includegraphics[width=0.49\linewidth, trim=0 0 -10 0]{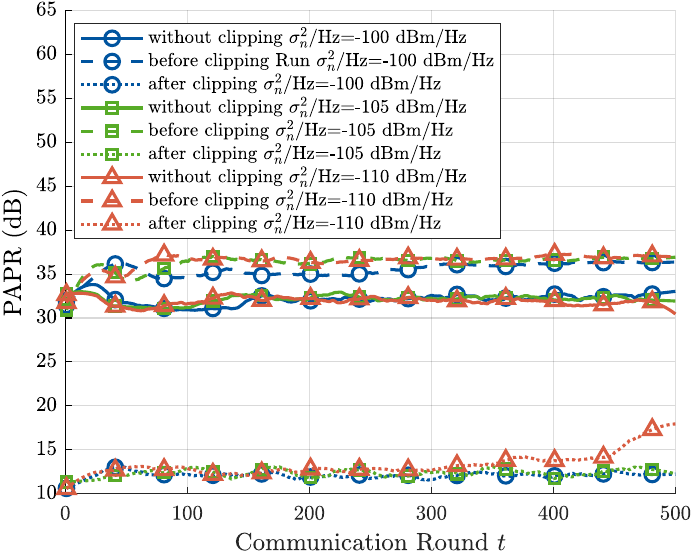}}
    \caption{Peak-to-average power ratio (PAPR) w.r.t. \acp{CR}.}
    \label{fig: papr}
\end{figure}

\Ac{PAPR} is shown in Fig.~\ref{fig: papr}. Without clipping, the PAPR is high (about \SI{35}{\decibel}), while after clipping and filtering, the PAPR remains close to \SI{15}{\decibel}. Note that the results without clipping and before clipping are different since PAPR depends on the actual values to be transmitted and clipping and filtering procedure changes the actual values to approximate the average of gradient~\eqref{eq: target OTA aggregation}. We also observe that single-carrier transmission can exhibit higher PAPR than multi-carrier OFDM transmission, which is unusual for digital modulation schemes. This stems from the nature of amplitude modulation (for the single-carrier case), where the power is directly related to the actual values to be transmitted, which can actually go up to ``infinity" if the gradient values can take arbitrarily large values.
The decrease in PAPR in the single-carrier case w.r.t. \acp{CR} may be due to that the model has been well trained, so the gradient value variance is smaller.

\begin{figure}[t]
    \centering
    \subfloat[Single-carrier\label{fig: tdma_error}]{
\includegraphics[width=0.49\linewidth]{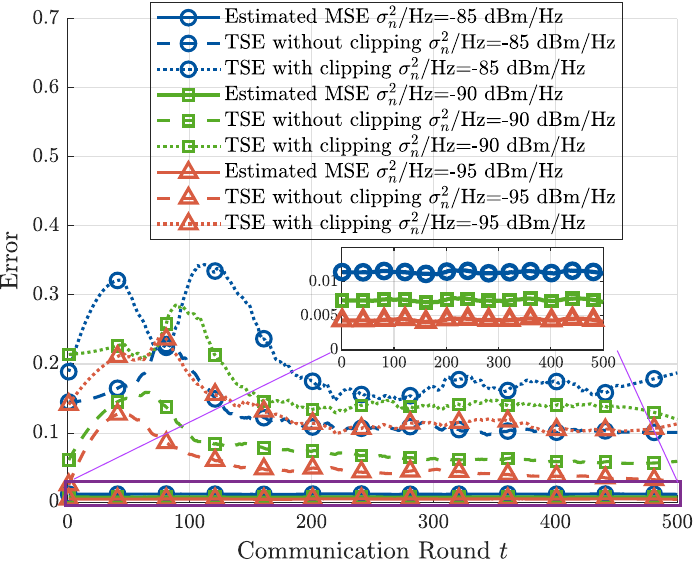}
    }
 \subfloat[Multi-carrier.\label{fig: OFDM_error}]{
\includegraphics[width=0.49\linewidth]{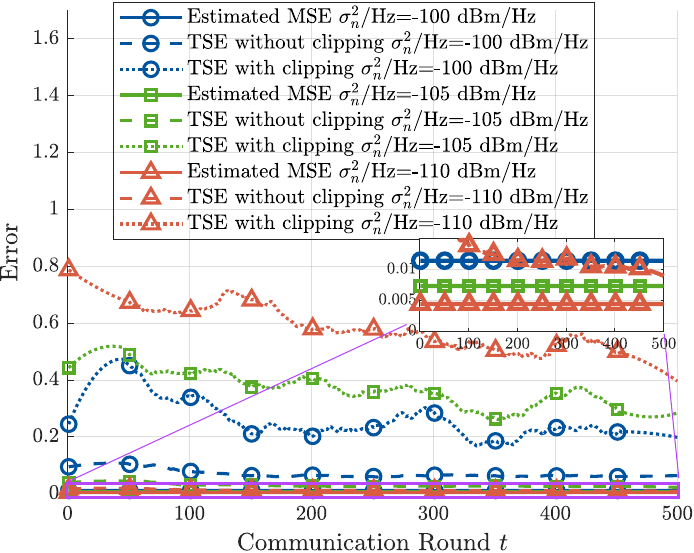}}
\vspace{-.1cm}
    \caption{Error of the approximated average gradient w.r.t. the actual average gradient. TSE denotes the \textit{True Square Error}; Estimated MSE is given by~\eqref{eq:MSE_TDMA} and~\eqref{eq: MSE_OFDM}.}
    \label{fig: error}
\end{figure}

The approximation error for~\eqref{eq: target OTA aggregation}, i.e. the MSE in~\eqref{eq:MSE_TDMA} and~\eqref{eq: MSE_OFDM}, is shown in Fig.~\ref{fig: error}. The true squared error (TSE) is defined as $\frac{1}{N}\sum_{n=1}^N\|\tilde{g}[n]-\overline{g}[n]\|^2$. Clipping increases the TSE relative to the non-clipping baseline, and the analytical MSE underestimates the TSE. Interestingly, for multi-carrier OFDM, the post-clipping TSE is larger when the noise power is lower. The reason is that, at low noise, the scaling factors~$\alpha_n$ take larger values, therefore amplifying the in-band distortion caused by \ac{ICF}, resulting in higher error.

\begin{figure}[t]
    \centering
    \subfloat[Single-carrier\label{fig: tdma_acc}]{
\includegraphics[width=0.49\linewidth, trim=0 0 -15 0]{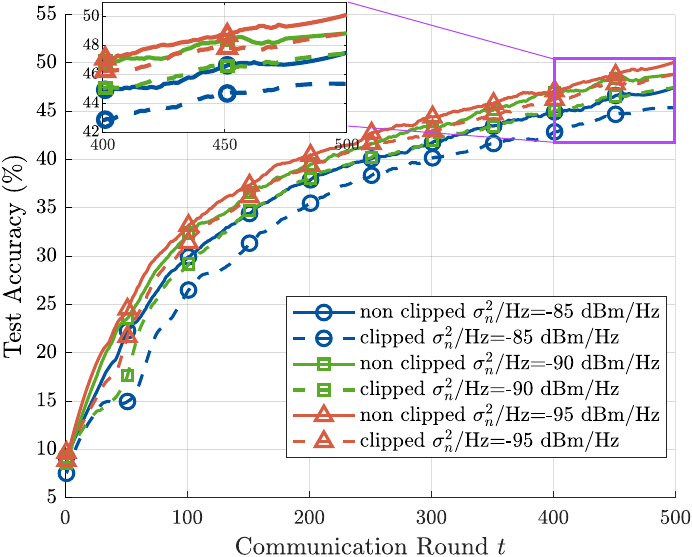}
    }
 \subfloat[Multi-carrier.\label{fig: OFDM_acc}]{
\includegraphics[width=0.49\linewidth, trim=0 0 -15 0]{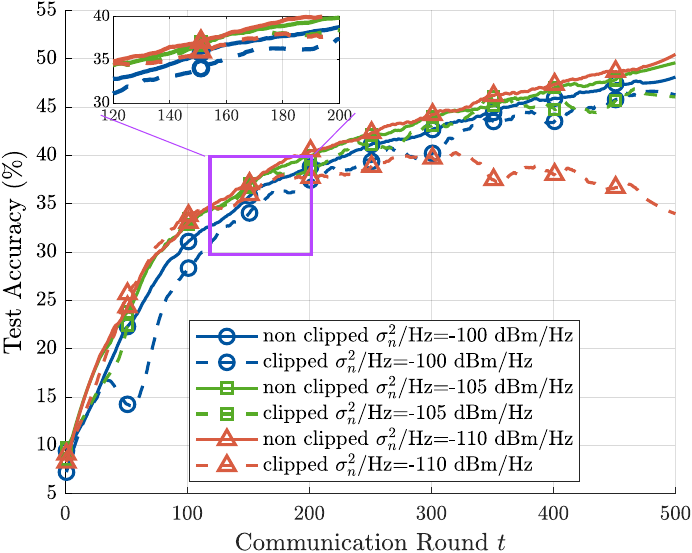}
}
    \caption{Test Accuracy w.r.t. \acp{CR}.}
    \label{fig: acc}
\end{figure}

\begin{figure}[t]
    \centering
    \subfloat[Single-carrier\label{fig: tdma_accdiff}]{
\includegraphics[width=0.49\linewidth, trim=0 0 -15 0]{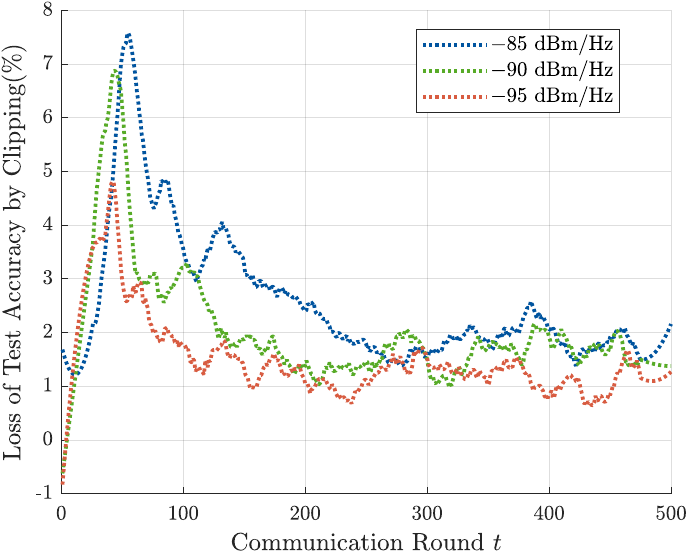}
    }
 \subfloat[Multi-carrier.\label{fig: OFDM_accdiff}]{
\includegraphics[width=0.49\linewidth, trim=0 0 -15 0]{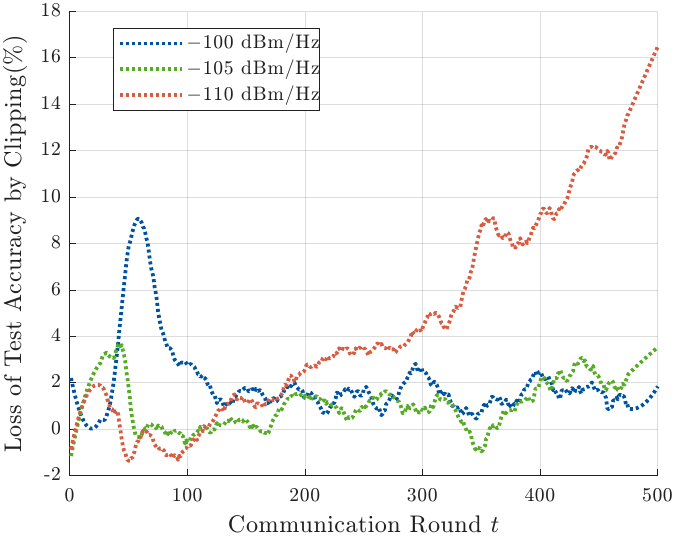}}
    \caption{Loss of test accuracy by imposing the peak power constraint (after clipping and filtering).}
    \label{fig: accdiff}
\end{figure}

The results on FL test accuracy are shown in Fig.~\ref{fig: acc}. For single-carrier (Fig.~\ref{fig: tdma_acc}), we observe consistent but small gaps between clipped and unclipped cases, with higher noise yielding lower accuracy overall. A noticeable accuracy drop appears at around 50 \acp{CR} for the clipped case. For multi-carrier OFDM (Fig.~\ref{fig: OFDM_acc}), as predicted by the TSE, the clipped case performs worse at lower noise; at $\sigma_n^2/\mathrm{Hz}=\,$\numdBm{-110}/Hz, training may even diverge. The corresponding accuracy loss due to clipping (considering the peak power constraint) is shown in Fig.~\ref{fig: accdiff}. At the later stage of training, the loss of test accuracy is about 1.5–2\%, while around 50 rounds it can peak at up to 8\% for the drop of accuracy mentioned previously.


\section{Conclusion}
\vspace{-.1cm}
This work highlights an overlooked practical issue in \ac{OTA}-\ac{FL}: the instantaneous peak power constraint imposed by the nonlinearity in power amplifiers. This is a practical issue in \ac{OTA}-\ac{FL} since amplitude-modulated analog transmission is by itself a potentially high \ac{PAPR} signal (depending on the distributions of values to be transmitted) and multi-carrier \ac{OTA}-OFDM would inherit the high PAPR issue of OFDM. 
As the first work that reveals this problem in AirComp-FL, in order to confirm that this is actually a problem to be solved, we applied the most classic PAPR reduction method of iterative amplitude clipping and frequency-domain filtering that constrains the signal within its PAPR limits without causing distortions to adjacent frequency bands. We observe that enforcing peak power constraint can indeed degrade aggregation and final accuracy, in some cases, the effects are marginal, and in some other cases (multi-carrier with low noise level), cause divergence. Counterintuitively, the impact is stronger at lower noise for multi-carrier OFDM because the scaling factor from optimization is higher and therefore the in-band distortion is too much amplified. This work should motivate further research in peak power constraint-aware power control and better PAPR reduction techniques for single-carrier and multi-carrier \ac{OTA}-\ac{FL} systems.

\bibliographystyle{IEEEtran}
\bibliography{biblio.bib}

\end{document}